\documentclass[aps,prb,twocolumn,superscriptaddress,showpacs]{revtex4-1}

\usepackage{bm}
\usepackage[colorlinks=true,linkcolor=blue]{hyperref}
\usepackage{graphicx}   
\usepackage{subfigure}  
\usepackage{amsmath}
\usepackage{amssymb}
\usepackage{gensymb}
\usepackage{soul}
\usepackage{ulem}

\newcommand{\vect}[1]{\,\overrightarrow{#1}}

\begin{document}

\title{Triangular array of iron-oxide nanoparticles:  A simulation study of intra- and inter-particle magnetism}

\author{B. Alkadour}
\affiliation{Department of Physics and Astronomy, University of Manitoba, Winnipeg, Manitoba, Canada R3T 2N2}

\author{B. W. Southern}%
\affiliation{Department of Physics and Astronomy, University of Manitoba, Winnipeg, Manitoba, Canada R3T 2N2}

\author{J. P. Whitehead}
\affiliation{Department of Physics and Astronomy, University of Manitoba, Winnipeg, Manitoba, Canada R3T 2N2}
\affiliation{Department of Physics and Physical Oceanography, Memorial University of Newfoundland, St. John's, Newfoundland, Canada A1B 3X7}

\author{J. van Lierop}
\affiliation{Department of Physics and Astronomy, University of Manitoba, Winnipeg, Manitoba, Canada R3T 2N2}

\begin{abstract}
A study of spherical maghemite nanoparticles on a two dimensional triangular array was carried out using a stochastic Landau-Lifshitz-Gilbert (sLLG) approach. The simulation method was first validated with a triangular array of simple dipoles, where results show the expected phase transition to a ferromagnetic state at a finite temperature.  The ground state exhibited a continuous degeneracy that was lifted by an order-from-disorder mechanism at infinitesimal temperatures with the appearance of a six-fold planar anisotropy.
The nanoparticle array consisted of 7.5~nm diameter maghemite spheres with bulk-like superexchange interactions between Fe-ions in the core, and weaker exchange between surface Fe-ions and a radial anisotropy. The triangular nanoparticle array ordered at the same reduced temperature as the simple dipole array, but exhibited different behaviour at low temperatures due to the surface anisotropy.  We find that the vacancies on the octahedral sites in the nanoparticles combine with the surface anisotropy to produce an effective random temperature-dependent  anisotropy for each particle.  This leads to a reduction in the net magnetization of the nanoparticle array at zero temperature compared to the simple dipole array.
\end{abstract}

\date{\today}

\maketitle

\section{Introduction}\label{sec:intro}

Nanomagnetic materials are used in a wide range of applications due to their versatile properties that depend on their shape and size.  This versatility is due to the complex nature of the magnetic structure of the individual nanoparticles, determined by interatomic exchange interactions between atoms and the single ion anisotropy, and the correlations between the nanoparticles which are determined by the long range dipolar interactions. Understanding the effect of each of these interactions and the subtle interplay between them is essential to the design and development of new nanoscale materials for applications, including those involving magnetic hyperthermia and contrast agents for magnetic resonance imaging\cite{yang}, spintronics\cite{wolf} and targeted drug delivery\cite{waha}.

For nanoparticles composed of transition metal oxides, surface effects (broken surface bonds and distorted coordination) give rise to strong inhomogeneities in both the single site anisotropy and the exchange parameters. As we show in this paper, the interplay between these interactions and the  vacant octahedral sites on maghemite nanoparticles play an important role in their magnetic behavior. While the precise nature of this surface anisotropy is somewhat elusive it can nevertheless be tuned by changing the surface to volume ratio of the nanoparticles, or by judicious doping\cite{Oberdick.2018,Skoropata.2017,Skoropata_book.2017}.  Such control over the magnetic structure of the transition metal nanoparticles provides a means of synthesizing magnetic nanoparticles with very specific properties.

In the case of nanoparticle assemblies the large magnetic moment of the individual nanoparticles can result in an enhanced, long range dipolar interaction between the individual nanoparticles.
 The impact of dipole interactions are of particular interest as it depends on the geometric arrangement (or lack-thereof) of the nanoparticles (e.g. into a superlattice/superstructure) rather than the chemical bonds and resulting exchange.  Manipulating the magnetic and wave-guide properties of assemblies of magnetic nanoparticles should be possible by changing the arrangement and spacing between the nanoparticles.  For example, while a triangular array of simple point dipoles orders ferromagnetically\cite{russier,Tomita}, magnetic order for a square array has anti-parallel stripes\cite{MacIssac,DeBell,russier}.   The structural sensitivity of low dimensional nanoparticle superlattices combined with the tuneability of the individual nanoparticles offer the potential to manipulate their magnetic and wave-guide properties in novel ways. To fully exploit this potential, it is essential to understand the often subtle and complex interplay between the long range correlations of the nanoparticles driven by the long range dipolar interaction between them and the magnetic properties of the individual nanoparticles.

In this paper we present a theoretical study of assemblies of single domain maghemite ($\gamma$-Fe$_2$O$_3$) spherical nanoparticles on a triangular array. The simulations studies are based on a hierarchical approach used previously to study a FCC superlattice of spherical maghemite nanoparticles\cite{pinning,alkdour2017}  in which the nanoparticles are treated at an atomistic level using the stochastic Landau-Lifshitz-Gilbert (sLLG) equations\cite{brown1963,llg} and the dipolar interactions between the nanoparticles  are incorporated through self consistent magnetic fields calculated based on a point dipole model with periodic boundary conditions. This approach is readily parallelized and hence can be applied to systems comprising 500 to 1,000 magnetic nanoparticles each of which comprises in excess of $10,000$ individual spins.

In this model, the nanoparticles comprise a core with bulk-like exchange interactions, and a surface layer with weaker exchange interactions and a single ion radial anisotropy. While in the temperature range of interest the core spins are aligned along a common axis, the surface magnetization is far more complex due to competition between the super exchange interactions and the surface radial anisotropy. This competition gives rise to a region of frustrated surface spins in a narrow band located at the magnetic equator. This combines with the random distribution of vacancies in the octahedral sites in maghemite to produce an  \textit{effective} anisotropy below the surface ordering temperature for each nanoparticle\cite{pinning}. Since this effective anisotropy is determined by the distribution of surface and the surface magnetization vacancies, its precise nature is unique to each individual nanoparticle and is strongly temperature dependent.


\section{Model and methodology:}\label{sec:model}

Maghemite has an inverse spinel structure in which each unit cell has 32 $\mathrm{O^{-2}}$,  8 $\mathrm{Fe^{+3}}$ occupying tetrahedral sites, and $16\times (5/6)$ $\mathrm{Fe^{+3}}$ occupying octahedral sites. One sixth of the octahedral sites are vacant to balance the charge.   Each nanoparticle in the lattice is described by a core-shell model that has an energy
\begin{equation}
\label{eq:NP-Energy}
E_\mathrm{NP}=  - \sideset{}{}\sum_{\langle ij\rangle} ( J_{i j} \hat{S}_{i}\cdot \hat{S}_{j} ) - K_{s}\sum_{i \in \mathrm{surface}} ( \hat{S}_{i}\cdot   \hat{n}_{i} )^{2},
\end{equation}
where $\hat{S}_{i}$ is a unit vector in the direction of the spin $i$, and $\hat{n}_{i}$ is a radially oriented unit vector.  The first term in Eqn.~(\ref{eq:NP-Energy}) describes the super-exchange interactions between the iron atoms that give rise to the ferrimagnetism.  The second term in Eqn.~(\ref{eq:NP-Energy}) describes the radial surface anisotropy ($K_s$) of the surface Fe spins.  Since the magnetic moment per Fe cation ($g_{s} S = 5 \mu_{B}$) is relatively large, we use a classical Heisenberg spin description.

To describe the long-range dipole interactions in the triangular array of these nanoparticles, we use the point dipole approximation.  In this approximation, the dipole field is calculated by representing each nanoparticle as a magnetic dipole at its centre where the magnetic moment of the dipole equals the magnetic moment of the corresponding nanoparticle. We construct an $L \times L$ triangular array with a lattice parameter $a$. Hence, the nanoparticles are located at sites $n_{1}\vect{a_{1}} +n_{2}\vect{a_{2}}$, where  $\vect{a_{1}}=a(1,0,0)$, $\vect{a_{2}}=a(1/2,\sqrt{3}/2,0)$, and $n_{1}$ and $n_{2}$ are integers between $1$ and $L$. To account for the long range interactions in the array, we impose periodic boundary conditions by using the Ewald summation method\cite{ewald-sqr}.  The dipolar energy of a triangular array of $L \times L$ dipoles is given by
\begin{equation}
\label{eq:dipoleEnergyi}
E_{d}=  \frac{1}{2} \sum_{k}^{L^{2}}\sum_{l}^{L^{2}} \sum_{\alpha,\beta=1}^{3} {\sigma}_{k}^{\alpha}\left(g_{k,l} \Gamma_{k,l}^{\alpha,\beta} \right) {\sigma}_{l}^{\beta},
\end{equation}
where $\Gamma_{k,l}^{\alpha,\beta}$ is an element of a $3\times3$ tensor that depends only on the relative position ($\vect{r}_{k,l}/a$) between the sites $k$ and $l$ in the array, $\sigma_{k}^{\alpha}$ is a Cartesian component of the normalized magnetic moment of the nanoparticle at site $k$, and $g_{k,l}=\mu_0 m_{k} m_{l}/4\pi a^{3}$ is the dipole interaction strength, where $m_{k}$ denotes the magnitude of the dipole moment of the nanoparticle at the site $k$, and $\mu_{0}$ is the vacuum permeability.  Therefore, the energy of a triangular array of the nanoparticles is
\begin{equation}
\label{eq:dipoleEnergy}
E=    \frac{1}{2} \sum_{k}^{L^{2}}\sum_{l}^{L^{2}} \sum_{\alpha,\nu=1}^{3} {\sigma}_{k}^{\alpha}\left(g_{k,l} \Gamma_{k,l}^{\alpha,\nu} \right) {\sigma}_{l}^{\nu} +  \sum_{q}^{L^{2}}   E_\mathrm{NP}^{q}  ,
\end{equation}
where $E^q_\mathrm{NP}$ is the energy due to the short-range interactions in the nanoparticle on the site $q$ in the array as defined in Eqn.~(\ref{eq:NP-Energy}).  The random vacancy distribution on the octahedral sites and the orientation of the maghemite unit cell are unique for each nanoparticle.

The simulations were carried out by using the RK4 scheme to integrate the sLLG equation\cite{llg}  given by
\begin{equation}
 \begin{split}
 &\frac{d\hat{S_i}}{dt}=
 -\frac{\gamma}{1+\alpha^{2}} [{\hat{S_i}} \times {\vect{H}}_{\mathrm{eff}}
 + \alpha \hat{S_i} \times (\hat{S_i} \times {\vect{H}}_{\mathrm{eff}}  )]\\
 & {\vect{H}}_{\mathrm{eff}} =-\frac{1}{\mu}\frac{\partial E}{\partial \hat{S_i  } } + \vect{H}_\mathrm{th},
  \end{split}
\end{equation}
where $\hat{S_i}$  is the normalized spin, $\gamma = 1.67\times 10^{11} \,(\mathrm{s\, T})^{-1}$ is the gyromagnetic ratio, $\alpha=0.5$ is the microscopic damping constant,  ${\vect{H}}_{\mathrm{eff}}$ is the effective magnetic field, $\mu$ is the local (spin or dipole) magnetic moment, and $\vect{H}_\mathrm{th}$ is the stochastic field to account for the thermal fluctuations.

The simulations of the arrays were carried out using the message passing interface, MPI,  where each nanoparticle was assigned to a single core-processor. The sLLG equation was integrated for each magnetic site in the nanoparticles in steps of $4 \times 10^{-4}\,\mathrm{tu}$, where $1\,\mathrm{tu} = 1.9 \times 10^{-11} \,\mathrm{s}$ and the dipole field at each nanoparticle was updated every 12 time-steps to reduce the communication time between processors.


\section{Dipole arrays}\label{sec:dipole}

\begin{figure}[h!]

\includegraphics[scale=0.8]{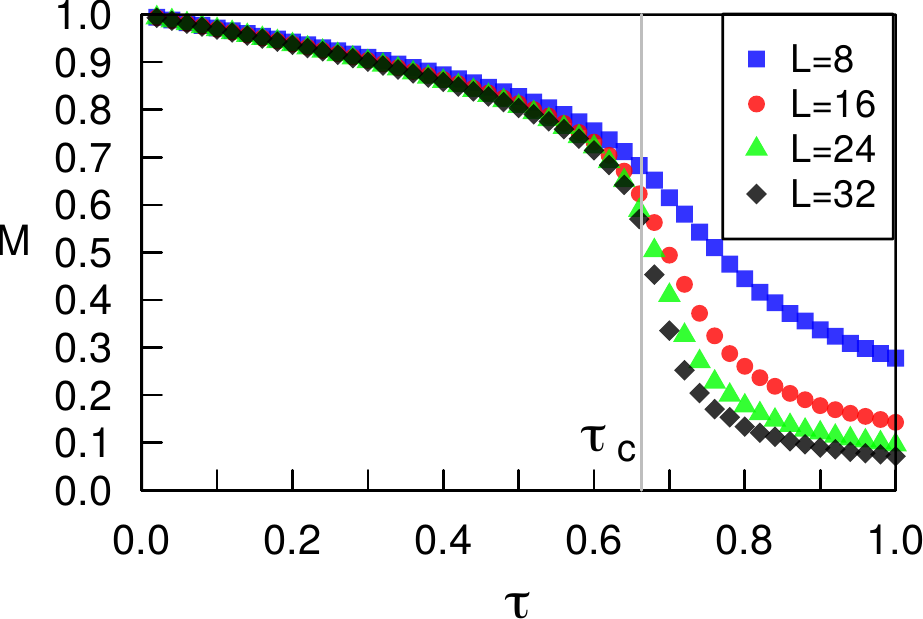}

\caption{(color online) The magnetization for different sizes of the triangular array of magnetic dipoles as a function of reduced temperature. \label{fig:M} }

\end{figure}

To validate our sLLG approach and to determine the nature of the magnetic order due to the dipole interactions in triangular arrays, we first performed simulations of simple dipoles in different sized triangular arrays.  Each ensemble consisted of 500 arrays of simple point dipoles, where the array size was $L$=8, 16, 24, or 32.  The magnetization and the energy were determined as a function of the reduced temperature $\tau =T k_B/g $ as the temperature was decreased from $\tau =1$ to 0 in steps of 0.02.  The arrays were left to equilibrate at each temperature step and the data were recorded after equilibration for statistical averaging.  Figure~\ref{fig:M} shows the magnetization as a function of temperature for the range of sizes; they all order ferromagnetically below a critical temperature.  Using finite size scaling of the Binder parameter (not shown), we found the critical temperature $\tau_c\simeq 0.663$, in good agreement with previous Monte-Carlo simulations by Tomita\cite{Tomita}.  Previous studies\cite{planar}  of dipole interactions in triangular arrays of planar spins showed a phase transition at $\tau_c \simeq 0.88$, which is slightly higher than our results for Heisenberg spins; out-of-plane fluctuations for the Heisenberg case can occur and suppresses $\tau_c$.


We found the ferromagnetic ground state to have energy-per-site, $E/g\simeq-2.757$, in agreement with previous studies\cite{bedanov,planar2}.  The ground state energy is independent of the direction of the net magnetization as long as it is in the plane of the array.  Since the dipole-dipole interactions acts like a planar anisotropy, our Heisenberg spin model and the planar spin model have identical ground state energies as pointed out by Rastelli~et~al.\cite{planar}.  Due to the finite value of the lattice size $L$, the net magnetization does not vanish above $\tau_c$ where the net magnetization of a fully random spin configuration would yield $M \propto 1/\sqrt{L^{2}}=1/L$.  We find that at low temperatures $M(0)-M(T)\propto T$.  This linearity is a characteristic of the classical Heisenberg model that we used to represent the dipoles (i.e. classical 3-dimensional dipoles)\cite{Yafet1,Yafet2,Evans}.

Our simulations identified that the net magnetization remains effectively in-plane below $\tau_c$.  For the planar spin system it has been shown that the ground state degeneracy is lifted and an order-from-disorder transition occurs at finite temperatures\cite{villain,henley,planar}.  As a result, a six-fold in-plane anisotropy arises in the planar spin system.  We have performed the same calculation for the Heisenberg case by expanding the free energy in terms of harmonic excitations for several values of $L$, and the array magnetization was found to prefer to be in the directions $\phi= 30^{\circ},90^{\circ},150^{\circ},210^{\circ},270^{\circ}$. We also found that the free energy barrier per site was dependent on the array size, as shown in Table~\ref{table:df}.


\begin{table}[h!]
\caption{The free energy barrier per site in units of $\tau$ for various array sizes as calculated from the expansion in term of the harmonic excitations.\label{table:df}}

 \begin{tabular}{|c|c|c|c|c|c|}
 \hline
  $L$ & 8 & 16 & 32 & 64 & 128 \\
 \hline
 $\Delta f/ \tau$	& 0.0344 & 0.0242 & 0.0205 & 0.0194 & 0.0190 \\
 \hline
\end{tabular}

\end{table}



\begin{figure}[t!]

\includegraphics[scale=0.9]{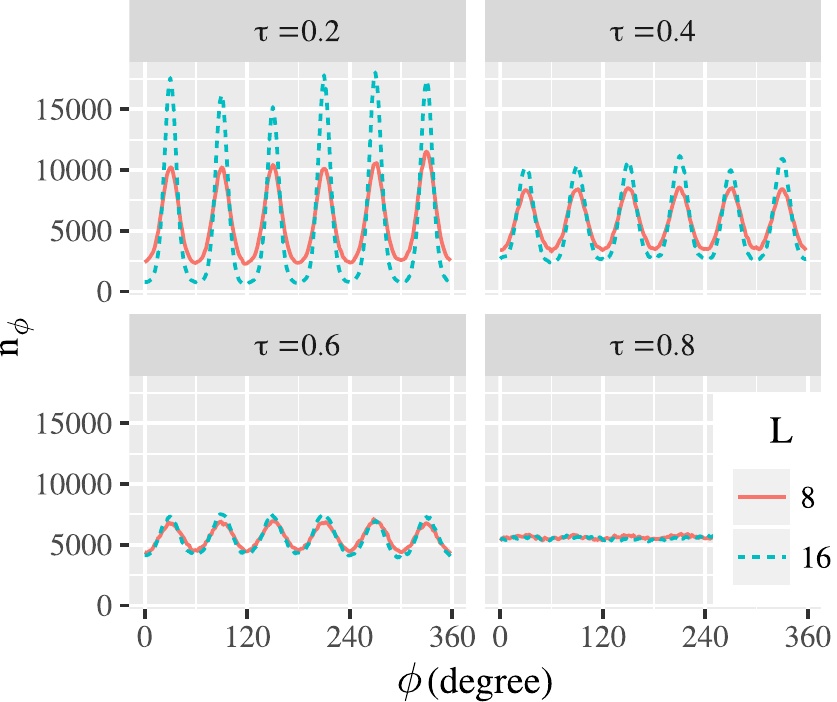}

\caption{(color online) Histogram of the angle between the net magnetization and $\vec{a}_{1}$ for array sizes $L=8$ (red) and $16$ (blue). \label{fig:theta} }

\end{figure}

\begin{figure}[t!]

\includegraphics[scale=0.9]{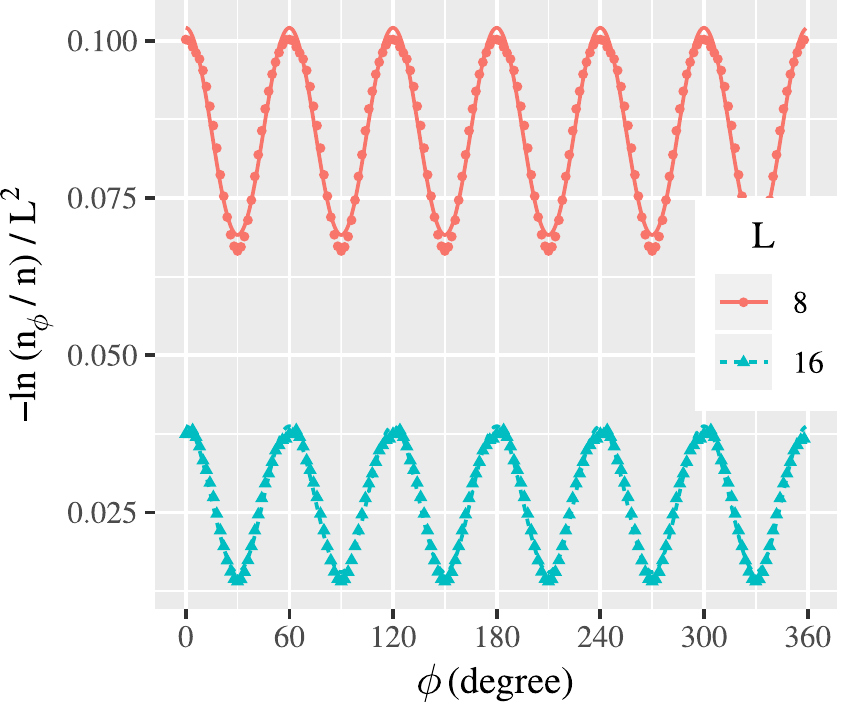}

\caption{(color online) $-\ln (n_{\phi}/n)/L^{2}$ as a function $\phi$ for ensembles of dipole arrays at $\tau=0.04$, $L=8$ (red) and 16 (blue). The points are the simulation results and the solid line is a fit with a cosine function as a guide to the eye.\label{fig:P}}

\end{figure}

We have also examined the behaviour of the close packed triangular dipole array for different sizes ($L$). For each temperature, we record 2000 measurements of the direction of the net magnetization of each array at equally spaced time intervals. We divide the planar angle $\phi$ between the net magnetization and $\vect{a}_{1}$ into sectors of 2 degrees and sum $n_{\phi}$, the number of times the net magnetization is found to be in each sector. Figure~\ref{fig:theta} shows $n_{\phi}$ at various reduced temperatures.  We find that the direction of the net magnetization exhibits a six-fold symmetry and prefers to be at a $30^{\circ}$ angle with respect to the primitive vectors.  At low temperatures, our arrays of Heisenberg spins behave like the arrays of planar spins\cite{planar}.   Given that the planar anisotropy  has a six-fold symmetry, we averaged over angles that were separated by $60^{o}$ intervals to improve the statistics.  The probability of the net magnetization being in a given sector is well approximated by $n_{\phi}/n$, where $n$ is the total number of measurements.  Since the Boltzmann factor $\exp{(-f_{\phi} L^{2}/ \tau)}\propto n_{\phi}/n$ where $f_{\phi}$ is the effective free energy per site in units of $g$, then $f_{\phi}/ \tau=-  \ln(n_{\phi}/n)/L^2+c$, where $c$ is a constant.  Figure~\ref{fig:P} shows $-\ln(n_{\phi}/n)/L^2$ as a function of $\phi$ at $\tau=0.04$ for sizes $L$=8 and 16. The effective free energy barrier per site can be estimated from the difference between the maximum and minimum values, given by $\Delta f_{\phi}/ \tau \simeq 0.0330$ for $L$=8, and $\Delta f_{\phi} / \tau \simeq 0.0232$ for $L$=16, in a good agreement with the predictions given in Table~\ref{table:df}. While the dipolar energy, given by Eqn.~(\ref{eq:dipoleEnergyi}), is not invariant under a uniform rotation of the spins, its ground state is nevertheless continuously degenerate. This property is common to a number of two dimensional dipolar systems\cite{deg}. This degeneracy can be removed by the presence of disorder resulting in long range magnetic order. The precise nature of the ordering depends on both the lattice structure and the form of the disorder. Such an effect is generally referred to as ``order from disorder" and can be induced by thermal fluctuations or structural disorder \cite{villain,henley,patchedjiev2005,patchedjiev2007}. While the net magnetization in our system remains in-plane below $\tau_c$, the individual dipoles are not necessarily in-plane.  This difference is important for nanoparticle arrays as we will see later.


\section{Maghemite Nanoparticles}\label{sec:maghemite}

\begin{figure}[b!]

\includegraphics[scale=0.75]{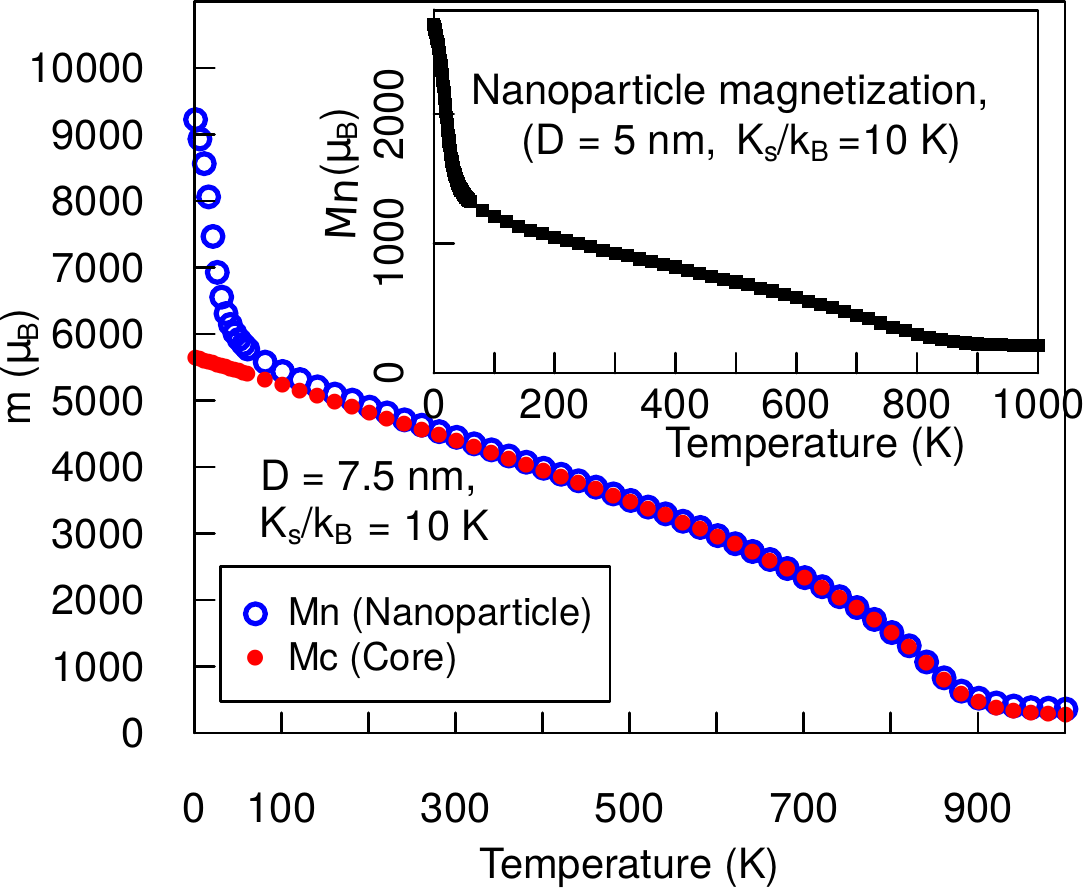}

\caption{(color online) The total and the core magnetization of a 7.5~nm diameter nanoparticle as a function of temperature. Previous results of $m_n(T)$\cite{pinning} for a 5~nm diameter nanoparticle are shown in the inset.
\label{fig:mNP}}

\end{figure}

The focus of this paper is on triangular arrays of maghemite spherical nanoparticles with a diameter of 7.5~nm (core diameter of 6.3~nm with the rest being the surface ``shell'');  the surface-to-volume ratio of  these nanoparticles is about 40\%, in keeping with typical experimental values.  We begin by examining the magnetism of non-interacting (no dipolar interactions) nanoparticles to determine their intrinsic spin moments and overall magnetization temperature dependencies.  The core is assumed to have bulk-like properties where the super-exchange between nearest neighbour, $ J_{TT},J_{OO},J_{OT}$, is -42, -17.2, and -56.2 K between tetrahedral-tetrahedral sites, octahedral-octahedral sites, and octahedral-tetrahedral sites, respectively \cite{pinning,kodamamaghemite,kodama2}.  These exchange constants were determined from magnetization and M\"ossbauer spectroscopy measurements on bulk maghemite using mean field theory\cite{sidorov1,sidorov,maghemiteTc,rado}.  However, mean field theory neglects fluctuations and usually overestimates the Curie temperature.  Monte Carlo and sLLG methods yield a lower Curie temperature and so we have scaled these exchange constants so that Curie temperature from sLLG yields reasonable agreement with the experimentally extrapolated values.  The ground state due to the super-exchange is ferrimagnetic where the spins of the octahedral sites are parallel to each other and anti-parallel to the spins on the tetrahedral sites, resulting in an average (effective) magnetization of 5/4$\mu_{B}$-per-magnetic-site. This is consistent with the experiments where the magnetization is found to be the result of two anti-parallel sublattices with a ratio of 0.6 between the number of spins on each sublattice.   The small magnetocrystalline anisotropy of the maghemite core is ignored for simplification (surface anisotropy effects will be discussed below).  Figure~\ref{fig:mNP} shows the core ($m_c(T)$) and the total magnetization ($m_n(T))$, of our 7.5~nm nanoparticle as a function of temperature obtained using the sLLG simulations.  The core spins order at 850~K, comparable to the ordering temperature of bulk maghemite\cite{maghemiteTc}, and for temperatures below 50~K the core spins are (effectively) aligned fully.


\begin{figure}[t!]

\includegraphics[scale=0.65]{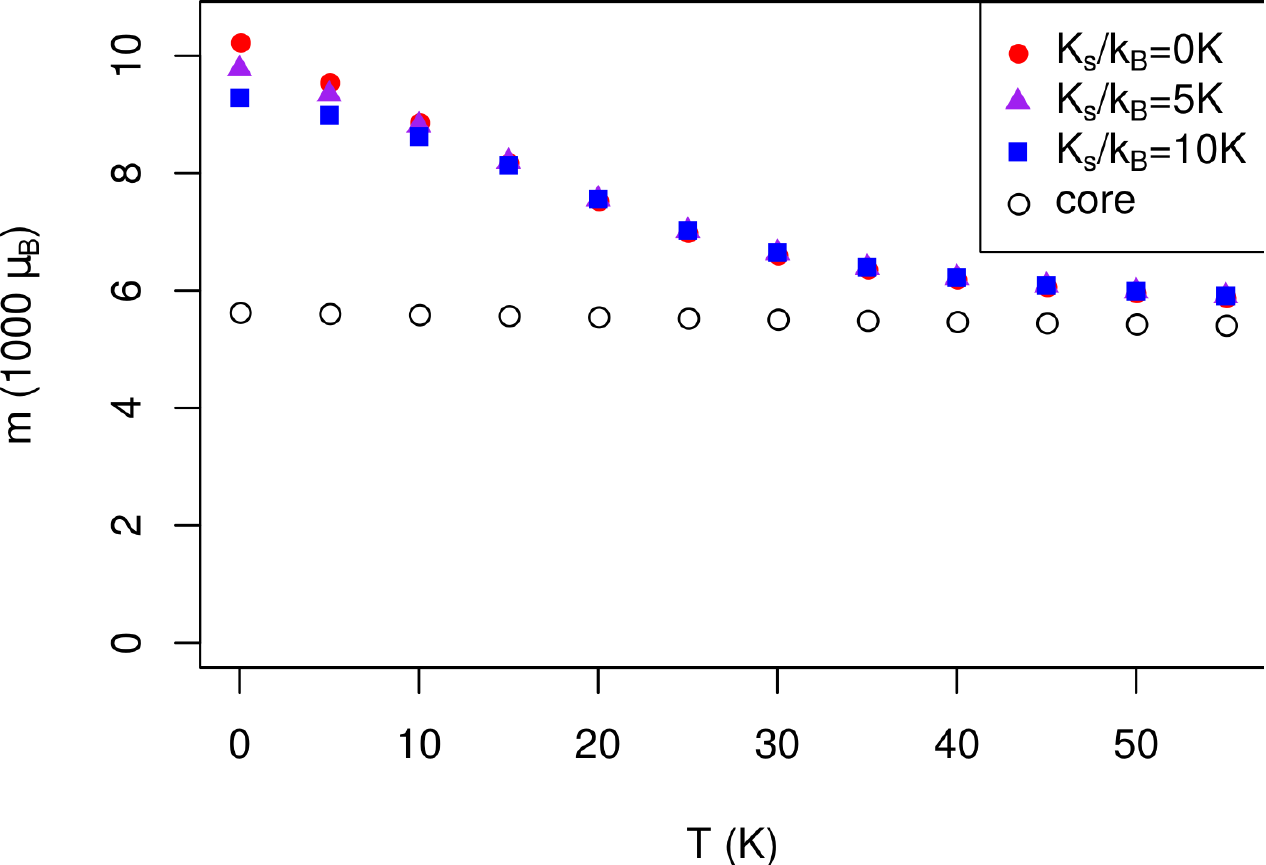}

\caption{(color online) The magnetic moment of individual nanoparticles as a function of temperature for various values of $K_s$. The open circles are the core magnetic moment which is saturated at low temperatures and is independent of $K_s$. \label{fig:KsNP}}

\end{figure}

Experimental results suggest that the surface spins order at a much lower temperature than the core.   To model this magnetism, we set the super-exchange between the surface spins to be 1/40 of the corresponding interactions between the core spins so that the surface spins start to order around 25~K. Below 25~K, the core spins are fully aligned, but the nanoparticle magnetization increases rapidly with decreasing temperature, as shown in Fig.~\ref{fig:mNP}.  This is in good agreement with previous simulations of 5~nm nanoparticles (inset of Fig.~\ref{fig:mNP})\cite{pinning} that were in agreement with previous Monte-Carlo simulations and experimental observations.  Figure~\ref{fig:KsNP} shows the magnetic moment of the 7.5~nm diameter nanoparticle for $K_s/k_B$=0, 5 and 10~K at low temperatures.  The core spins are fully aligned below 60~K, and below 15~K the surface spins are relatively well ordered.  We find that the nanoparticle's magnetization decreases with increasing $K_s$, suggesting that the surface radial anisotropy reduces the alignment between surface spins. 


\section{Triangular Nanoparticle Array}\label{sec:NParray}

To study the interplay that results from the surface anisotropy and dipole interactions between nanoparticles in a triangular array, simulations of an array size $L$=24 and lattice parameter $a$=7.5~nm were carried out. The  dipole-dipole interactions between nanoparticles were dealt with as discussed above.  The array was cooled from 55~K to 0 in steps of 5~K, and the simulation was repeated for radial anisotropy constants $K_s/k_B$=0, 5, and 10~K.

\begin{figure}[b!]

\includegraphics[scale=0.6]{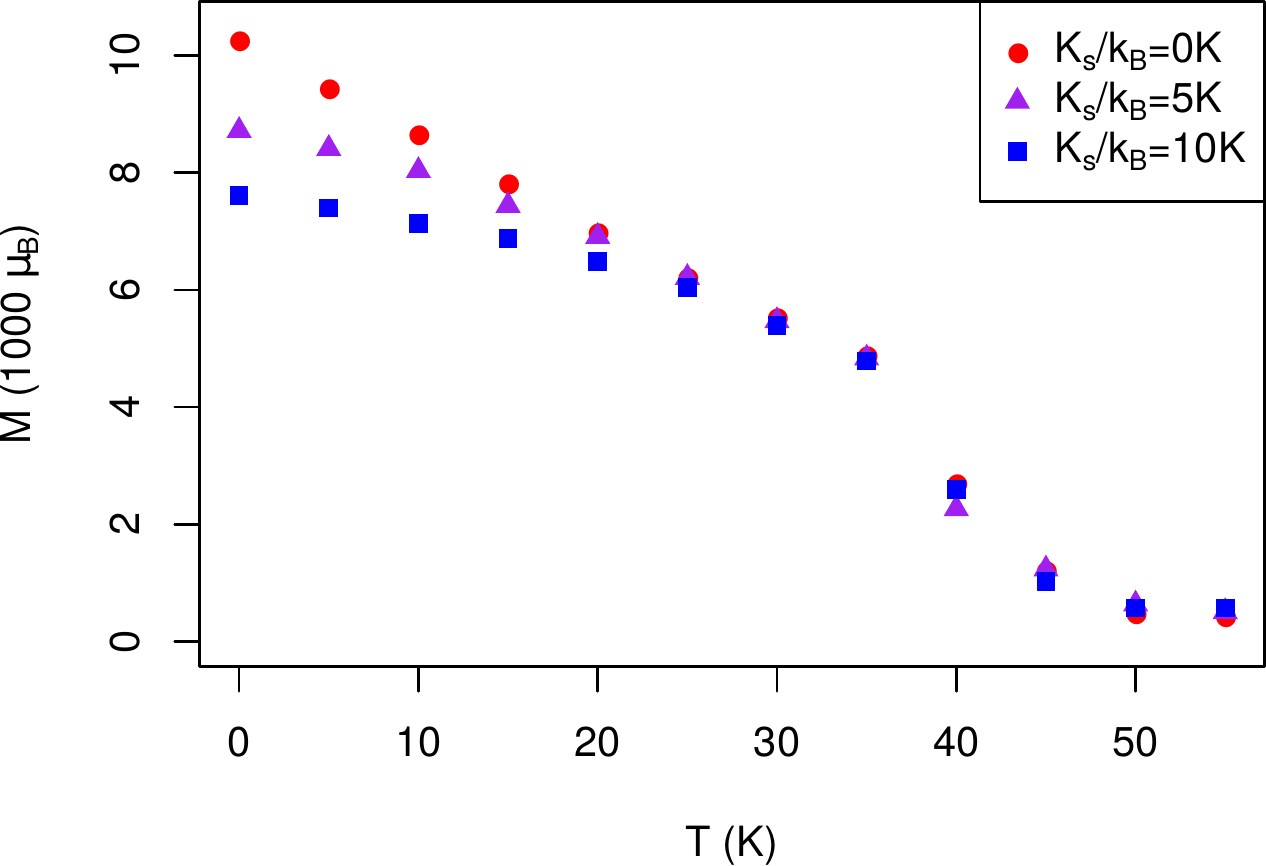}

\caption{(color online) The magnetization per nanoparticle of nanoparticle triangular arrays with L= 24 as a function of temperature for different values of $K_s$. \label{fig:MvsT-NP}}

\end{figure}

Figure~\ref{fig:MvsT-NP} presents the temperature dependence of the average magnitude of the array magnetization, $M(T)$, per nanoparticle.  While the magnetization of each individual nanoparticle was relatively large at 45~K (due to the ordering of core spins as shown in Fig.~\ref{fig:KsNP}),the net magnetization of the triangular array per nanoparticle is significantly less due to the lack of alignment of the nanoparticle magnetizations.


Since the nanoparticle magnetization, $m(T)$, increases significantly with cooling below 25~K (Fig.~\ref{fig:KsNP}) the strength of the dipole interactions between the nanoparticles, $g(T)=\mu_0 [m(T)]^{2}/4\pi a^{3}$, is strongly temperature dependent.  Due to finite size effects, the simple dipole array discussed previously with $L=24$ started to order at a reduced temperature $\tau \simeq 0.8$, higher than $\tau_c$.  For the nanoparticle array, this corresponds to
\begin{equation}\label{eq:tc}
\begin{split}
& \tau= T k_B /g(T) =0.8 \Longrightarrow  T \simeq (1.18 \times 10^{-6}) m(T)^2\\
& T \simeq  44 ~\mathrm{K},\\
& m(T=44\mathrm{K})=6100~\mu_B
\end{split}
\end{equation}

\begin{figure}[b!]

\includegraphics[scale=0.65]{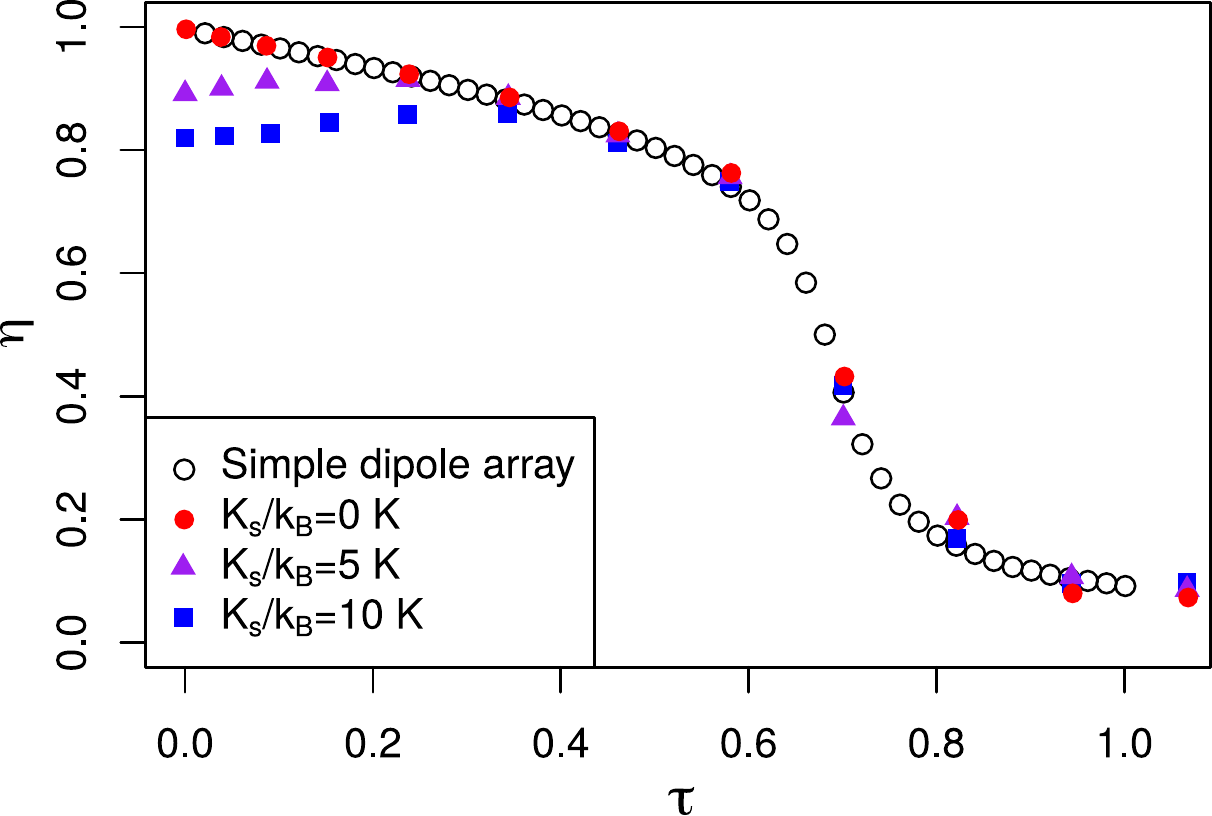}

\caption{(color online) The order parameter as a function of the reduced temperature for dipole and nanoparticle arrays with $L=24$. \label{fig:OT}}

\end{figure}

For temperatures below $25\ \mathrm{K}$, Fig.~\ref{fig:MvsT-NP} shows $M(T)$ decreasing with increasing $K_s$. While some of this decrease may be attributed to the decrease in the surface magnetization with increasing $K_s$, this is not the whole story. To understand this, we define an order parameter $\eta = M(T)/m(T)$ and a reduced temperature $\tau = T k_B/g(T)$. Plotting the order parameter $\eta$ as a function of the reduced temperature $\tau$, as shown in Fig.~\ref{fig:OT}, removes the dependence of the data on the net magnetic moment of the nanoparticles. The plots of the rescaled data shown in Fig.~\ref{fig:OT} show that the nanoparticles with no surface anisotropy ($K_s$=0) behave as simple dipoles. For $K_s \neq 0$, the order parameter $\eta(\tau)$ falls below the curve of the simple dipoles at low temperature ($\tau < 0.22$ for $K_s/k_B = 5\,\mathrm{K}$ and $\tau < 0.35$ for $K_s/k_B = 10\,\mathrm{K}$.


\begin{figure}[t!]

\includegraphics[scale=0.07]{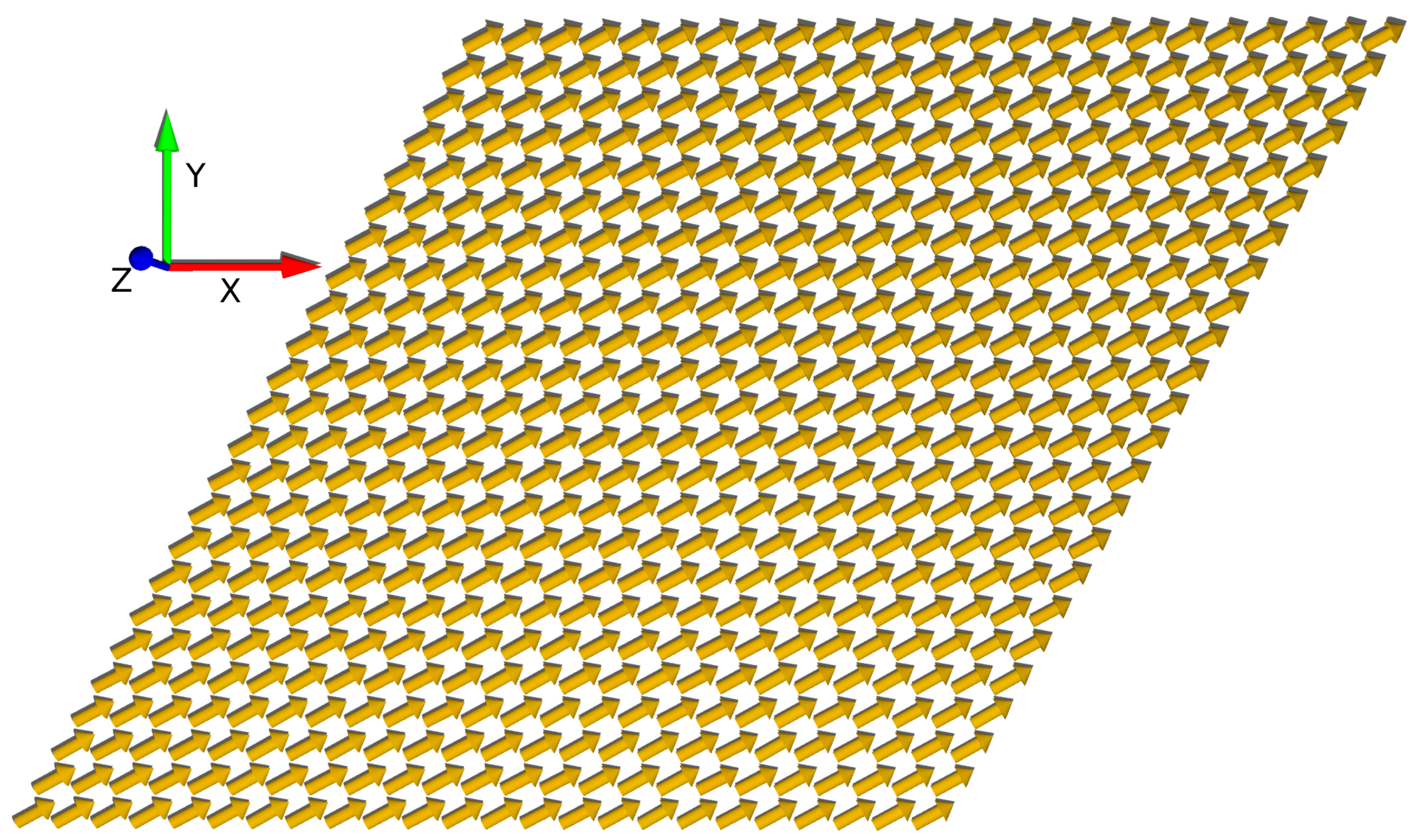}

\caption{(color online) The magnetic configuration of nanoparticle arrays with $K_s=0$ and $L=24$ at $T=0$. Each arrow represents the magnetization of one nanoparticle. Each arrow represents the magnetization of one nanoparticle. The directions of the arrows are color-coded using a composition of red, green, and blue colors to represents the x-,y- and z-components (see the main axes on the top left). The golden color of the arrows is the result of mixing red (x-component) with some green (y-component). \label{fig:Ks0T0}}

\end{figure}



\begin{figure}[t!]

\subfigure[\ ]{\includegraphics[scale=0.085]{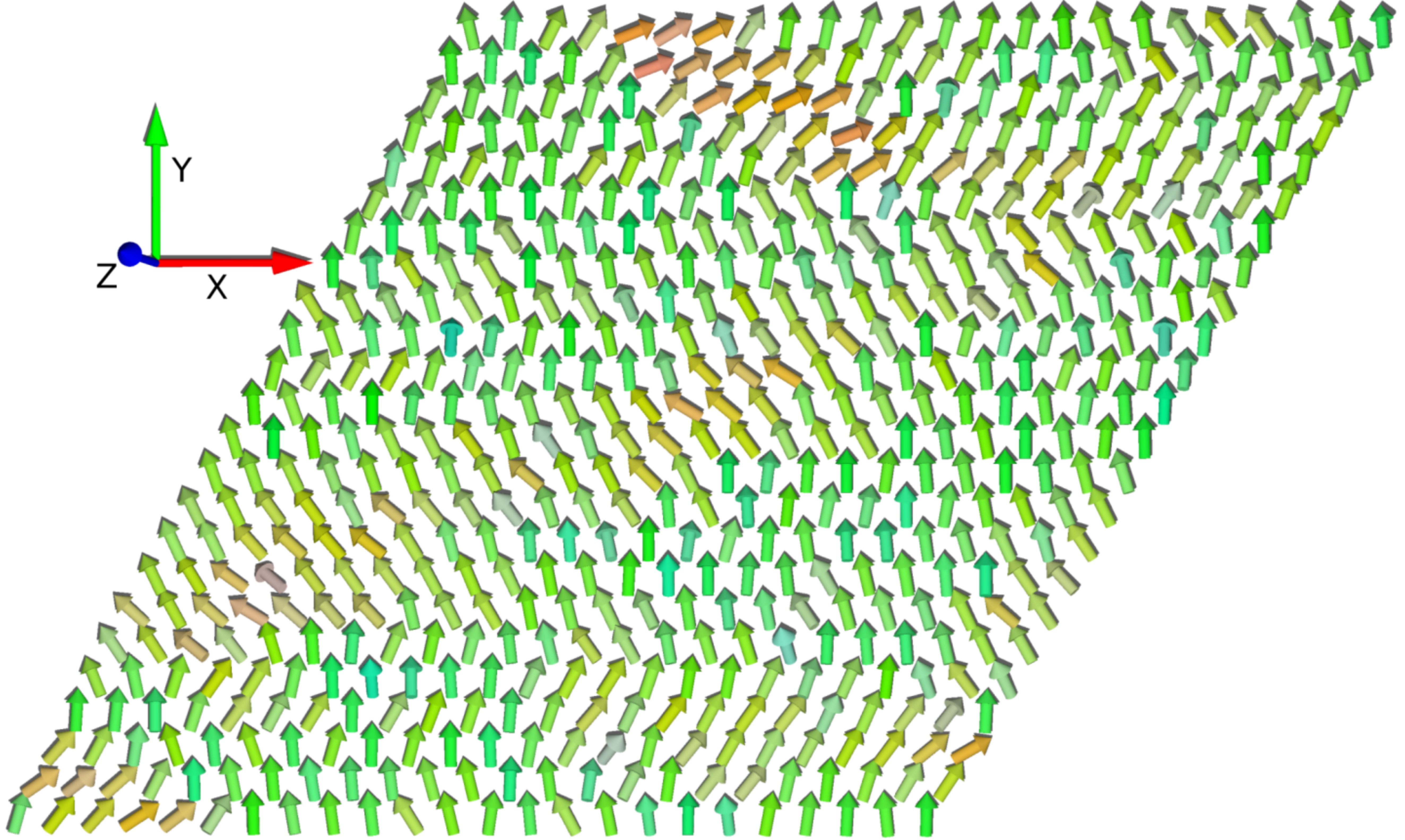}}
\subfigure[\ ]{\includegraphics[scale=0.085]{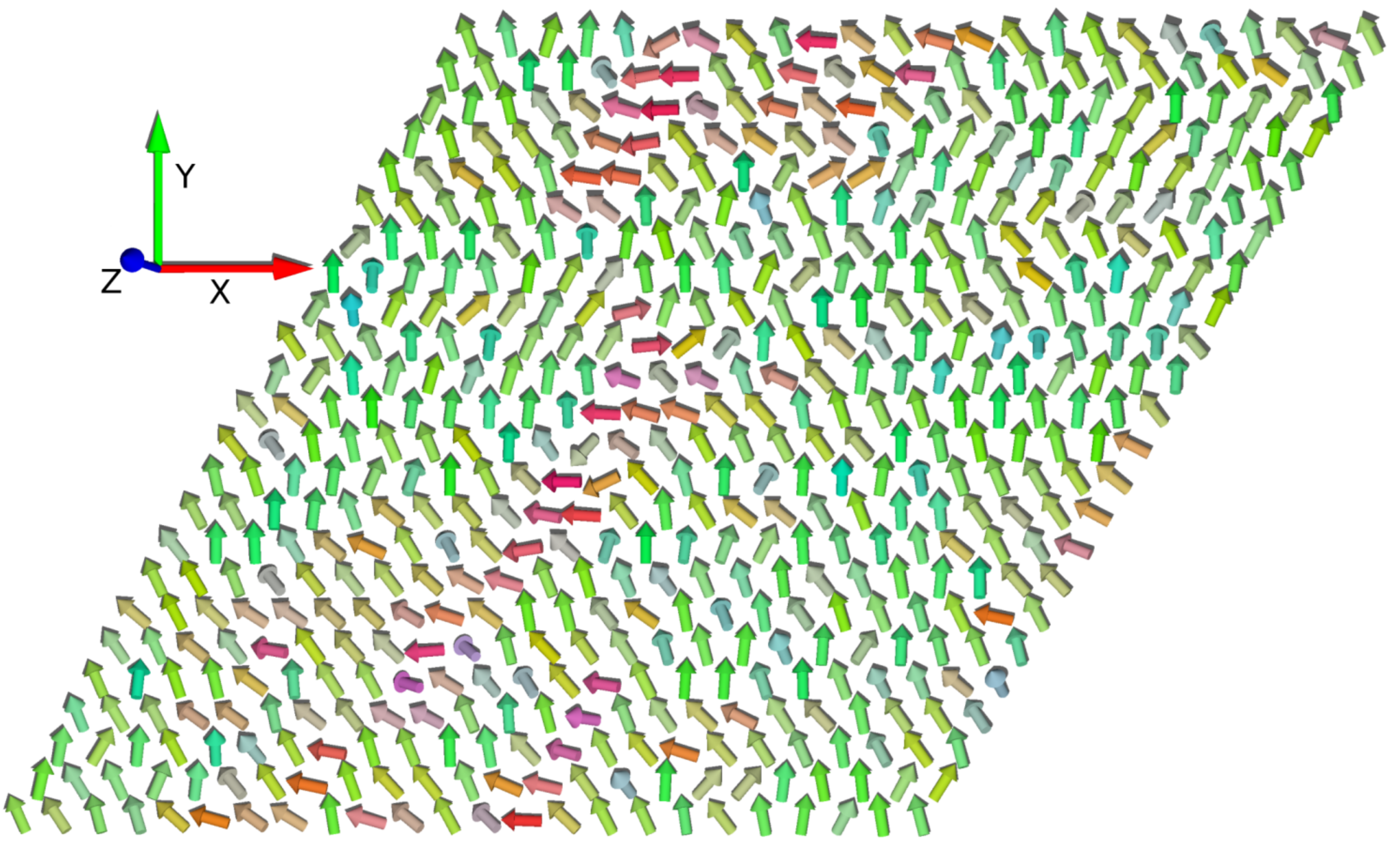}}

\caption{(color online) The magnetic configuration of nanoparticle triangular arrays of size $L$=24 with (a) $K_s/k_B$=5~K and (b) $K_s/k_B$=10~K  at $T=0$.  Each arrow represents the magnetization of one nanoparticle. The color coding scheme is the same as in Fig.~\ref{fig:Ks0T0}  \label{fig:ArrayConfig}}

\end{figure}


The effect of the surface radial anisotropy on $\eta$ can be illustrated by comparing the magnetic configuration of an array of nanoparticles for $K_s=0$ at $T=0$ ($\tau$=0) as shown in Fig.~\ref{fig:Ks0T0}  (each arrow represents the magnetization of one nanoparticle) with the corresponding magnetic configuration of the same array  for $K_s/k_B = 5\,\mathrm{K}$ and $10\,\mathrm{K}$ as shown in Fig.~\ref{fig:ArrayConfig}. Each of the configurations was obtained after cooling the array from 55 K to 0. In the case of $K_s=0$, cooling the array relaxes to a ferromagnetic state in which the nanoparticles are perfectly aligned ferromagnetically along one of the six preferred directions determined by the thermal fluctuations, as discussed in Section~\ref{sec:dipole}, giving an order parameter $\eta = 1$ for $\tau=0$. The corresponding magnetic configurations for the case of $K_s\neq 0$ show a much more complex structure with out-of-plane components in addition to an in-plane disorder that results in the formation of magnetic domains. These results are similar to those reported by Russier\cite{russier} who studied triangular arrays of nanoparticles that had no internal structure but a random uniaxial anisotropy at 0~K, and he observed a similar domain configuration. The inter-particle disorder in arrays of maghemite nanoparticles with non-zero radial anisotropy suggests that an \textit{effective} anisotropy arises for each nanoparticle. This is consistent with previous atomistic studies on FCC superlattice arrays of maghemite nanoparticles\cite{alkdour2017}. The nature and origin of this anisotropy is discussed in the following section.


\section{The Effective Anisotropy}\label{sec:effAnisotropy}

\begin{figure}[t!]

\includegraphics[scale=0.7]{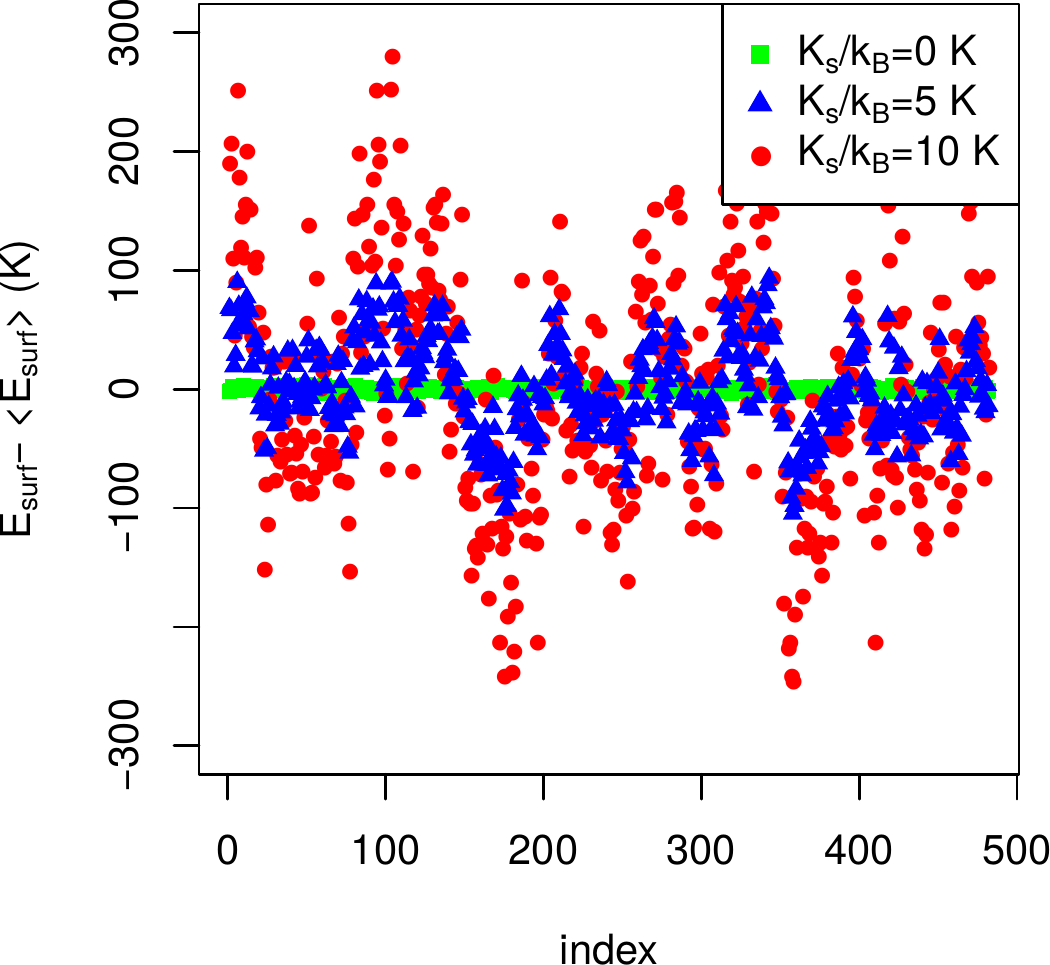}

\caption{(color online) The change in the surface energy plotted as a function of the index of the direction of the core magnetic moment calculated for a single nanoparticle. \label{fig:EvsK}}

\end{figure}


While the existence of an \textit{effective} anisotropy in the individual nanoparticles at low temperatures that is dependent on the radial anisotropy of the surface spins would appear to account for the reduction in the order parameter $\eta$ at low temperatures, the mechanism by which radial anisotropy associated with the surface spins can give rise to such a term is not obvious. To understand the origin and character of this \textit{effective} anisotropy we first note that for temperatures below which the surface spins begin to order, the magnetization of the nanoparticle core is essentially saturated and that the core Fe ions have no single site anisotropy. This implies that the effective anisotropy of the nanoparticles originates from the surface and its exchange with the fully saturated core. To illustrate this, we consider a single nanoparticle in which  the magnetic moment of the core spins are assumed to be fully saturated and to be in a fixed orientation, denoted by the unit vector $\hat\sigma$. Using the standard Monte Carlo method, the surface spins are given random directions and then allowed to equilibrate at a given temperature followed by a calculation of the total energy and magnetization of the surface spins for each of the core magnetization directions. This calculation is performed for $N$ values of $\hat\sigma$ which we denote by the set $\{\hat\sigma_i\} = (\hat\sigma_1,\hat\sigma_2 \dots \hat\sigma_N)$ and each of the data represented by a point on a unit sphere.

The surface energy is calculated as a function of the index $i$ for three values of $K_s/k_B=0\,\mathrm{K},\,5\,\mathrm{K}\,\mathrm{and}\,10\,\mathrm{K}$ at temperature of $2\,\mathrm{K}$ as shown in Fig.~\ref{fig:EvsK}. For $K_s=0$ there is no variation of the surface energy of the nanoparticle  due to the rotation of the core spins. This is consistent with the result in Fig.~\ref{fig:OT} which shows that the dependence of the rescaled order parameter $\eta$ on the reduced temperature $\tau$ agrees closely to the corresponding result for point dipoles. This is in sharp contrast to the case of finite values of $K_s$ for which  the surface energy of the nanoparticle shows a strong variation as a function of the index $i$. This behaviour suggests that the preferred orientation of the nanoparticle magnetization is determined by the variation in the surface energy with respect to the orientation of the core magnetization, $\hat\sigma_i$, and can be thought of as an effective anisotropy. 

\begin{figure}[t!]
\includegraphics[scale=0.5]{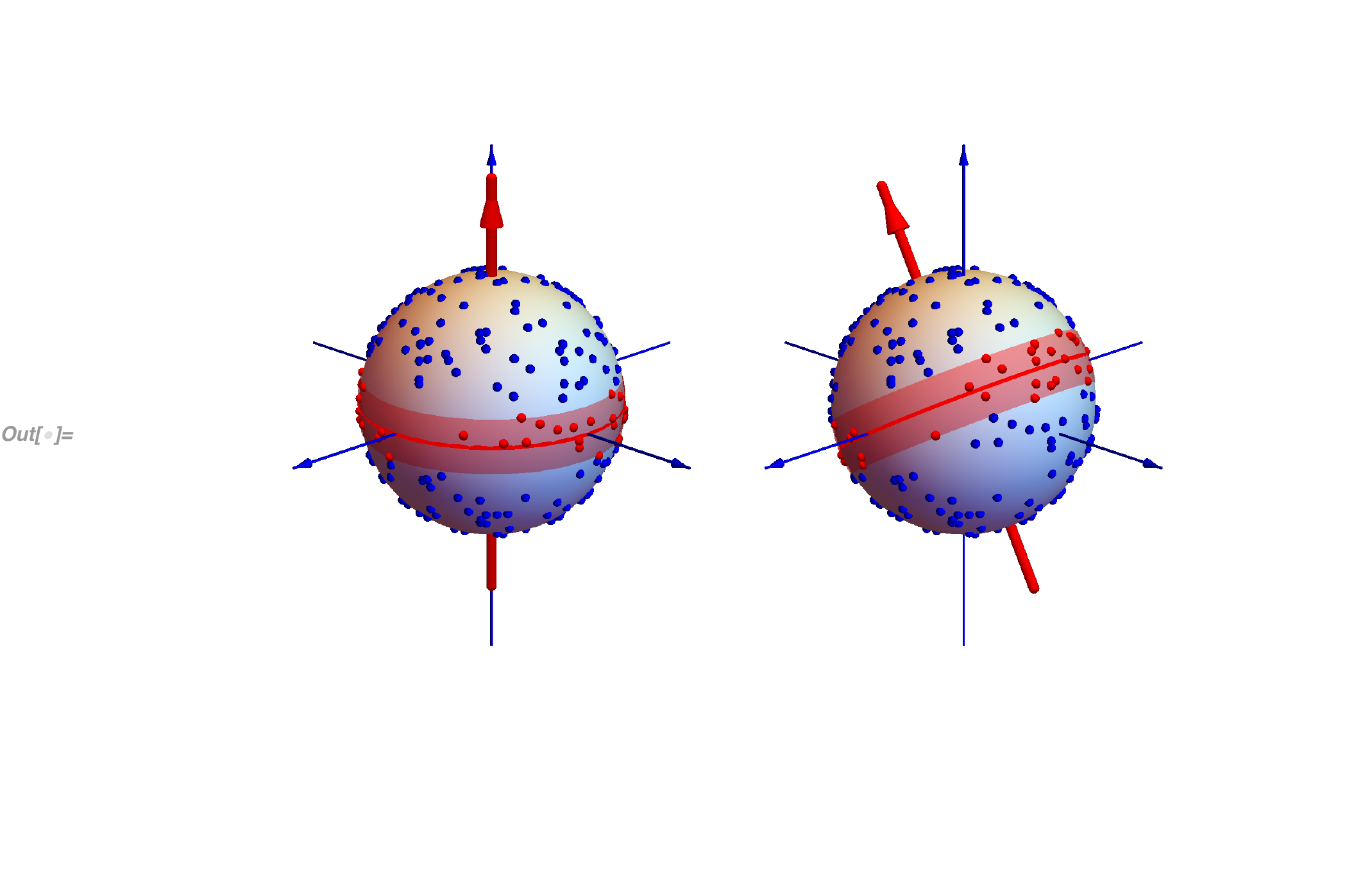}
\caption{(color online) The arrow indicates the direction of the core magnetization, the red line indicates the magnetic equator and the red
band indicates the region in which the number of vacancies on octahedral sites are enumerated. }\label{fig:vacanciesT}

\end{figure}

\begin{figure}[t!]

\includegraphics[scale=0.7]{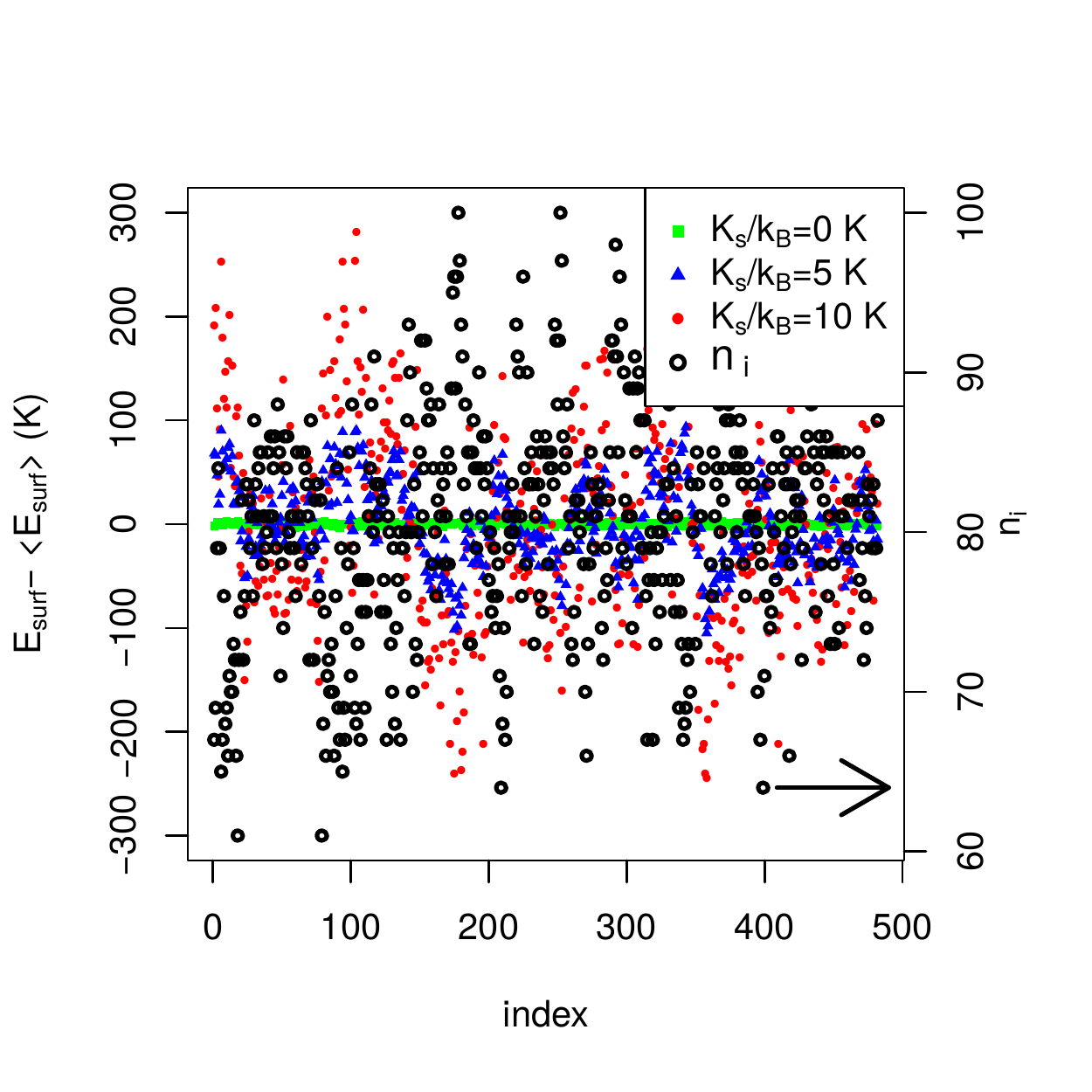}

\caption{ (color online) The change in the surface energy (the same data that are shown in Fig.~\ref{fig:EvsK}) is plotted along with the number of vacancies on the octahedral sites in the equatorial region for each direction of the core magnetization for the same nanoparticle.\label{fig:EvsK1}}

\end{figure}

To explore the origin of this effective anisotropy, we begin by noting that one sixth of the octahedral sites are vacant and that these vacant sites are randomly distributed throughout the nanoparticle. Due to statistical variation the vacancies in the nanoparticles will not be distributed uniformly. This is clearly seen in Fig.~\ref{fig:vacanciesT} which show the location of the surface vacancies projected onto the surface of the unit sphere for a typical nanoparticle. Of particular interest is the number of surface vacancies located in the narrow band of octahedral sites close to the equatorial plane perpendicular to $\hat \sigma$. Figures~\ref{fig:vacanciesT} (a) and (b) show the surface vacancies that are located within a band of width $\pm 11.5\degree$  for two values of $\hat\sigma$ indicated by the red arrows.  Denoting by $n_i=n(\hat\sigma_i)$ the number of surface vacancies that are located within this equatorial band of octahedral states for each of the core orientations $\{\hat\sigma_i\}$, Fig.~\ref{fig:EvsK1} plots $n_i$ as a function of the index $i$ together with the corresponding value for the energies $E_i=E(\hat\sigma_i)$ shown in Fig.~\ref{fig:EvsK}.  The data presented in Fig.~\ref{fig:EvsK1} illustrates two important features. The first is simply the variation in $n_i$ and the second is the fact that $n_i$ appears to be anti-correlated to the energy $E_i$. Using the surface energy data for $K_s/k_B = 10\,\mathrm{K}$ we have evaluated a Pearson's correlation $\frac{\sum_i  \Delta n_i\Delta E_i}{\sqrt{ (\sum_i \Delta n_i^2 )(\sum_i \Delta E_i ^2)}} =-0.68$ where $\Delta n_i = n_i - \langle n_i \rangle$ and $\Delta E_i= E_i - \langle E_i \rangle$ respectively.


\begin{figure}[t!]

\subfigure[\ ]{\includegraphics[scale=0.275]{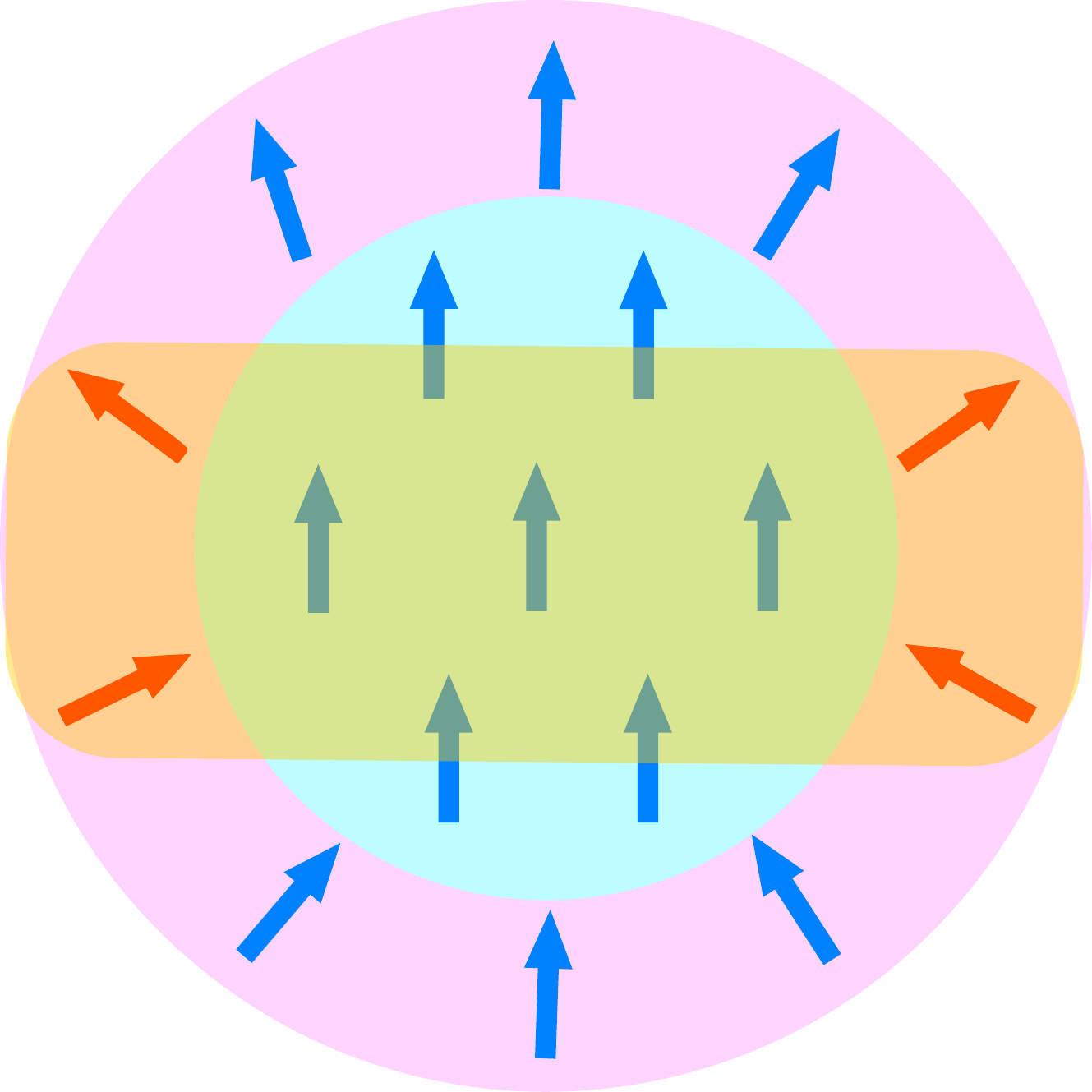}}
\subfigure[\ ]{\includegraphics[scale=0.275]{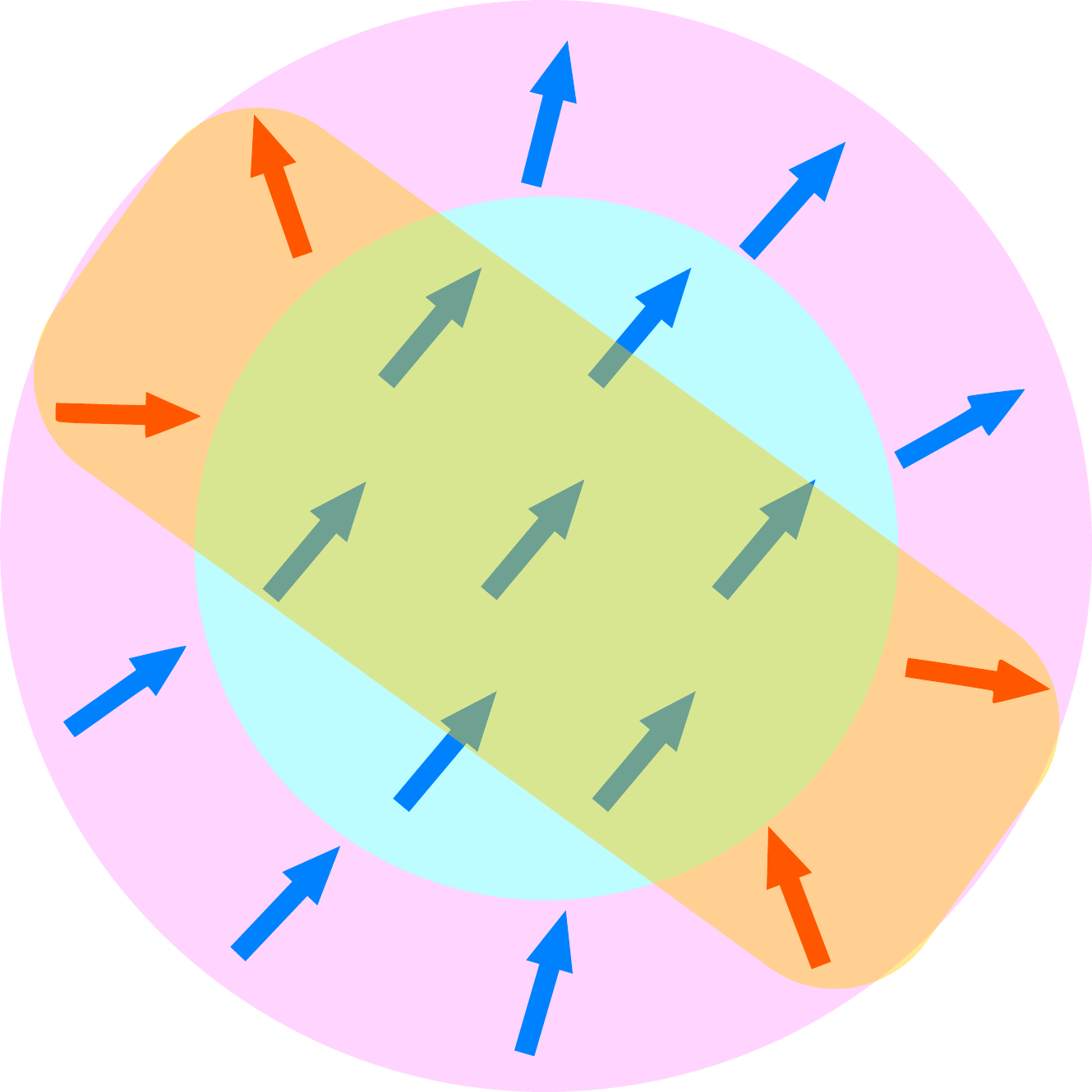}}

\subfigure[\ ]{\includegraphics[scale=0.275]{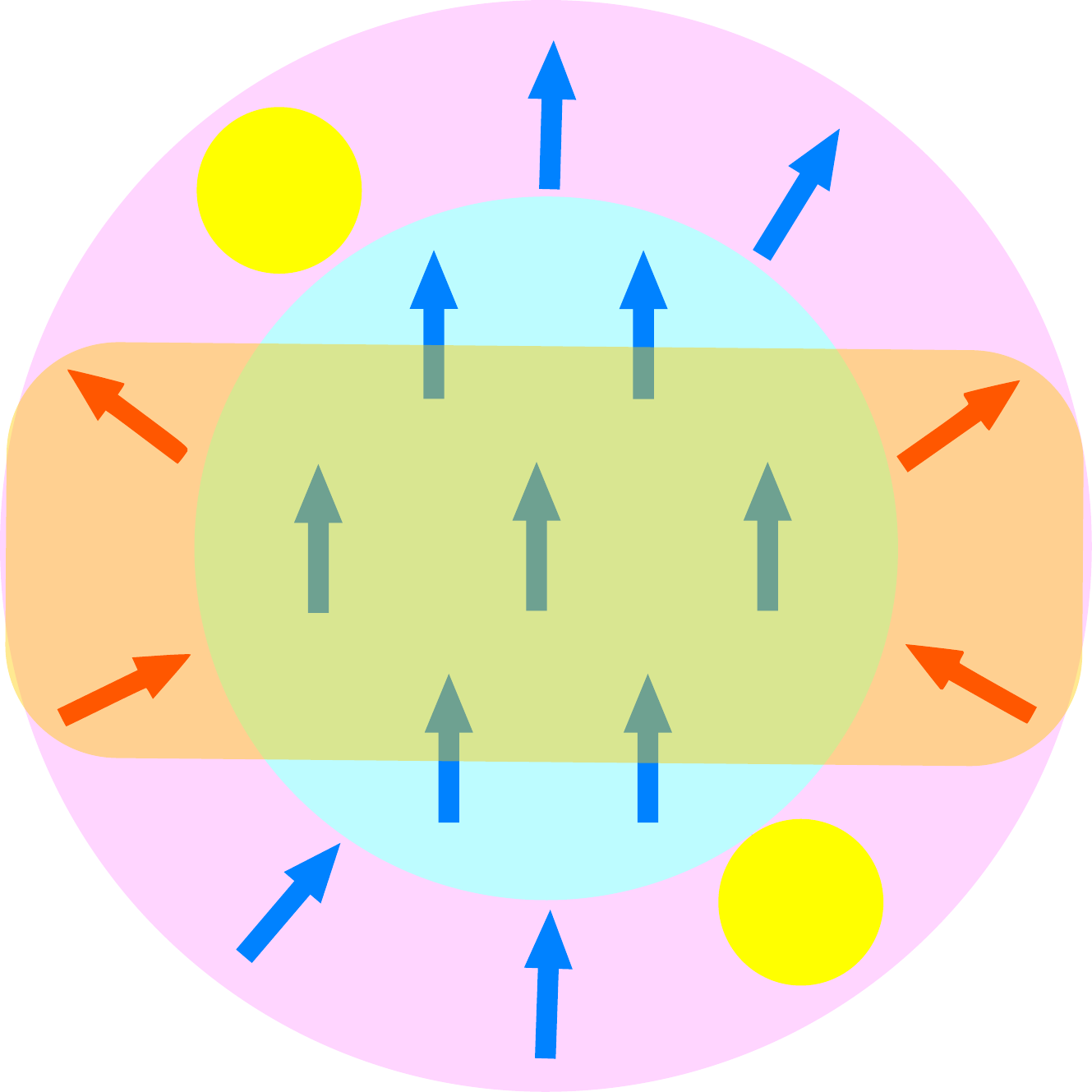}}
\subfigure[\ ]{\includegraphics[scale=0.275]{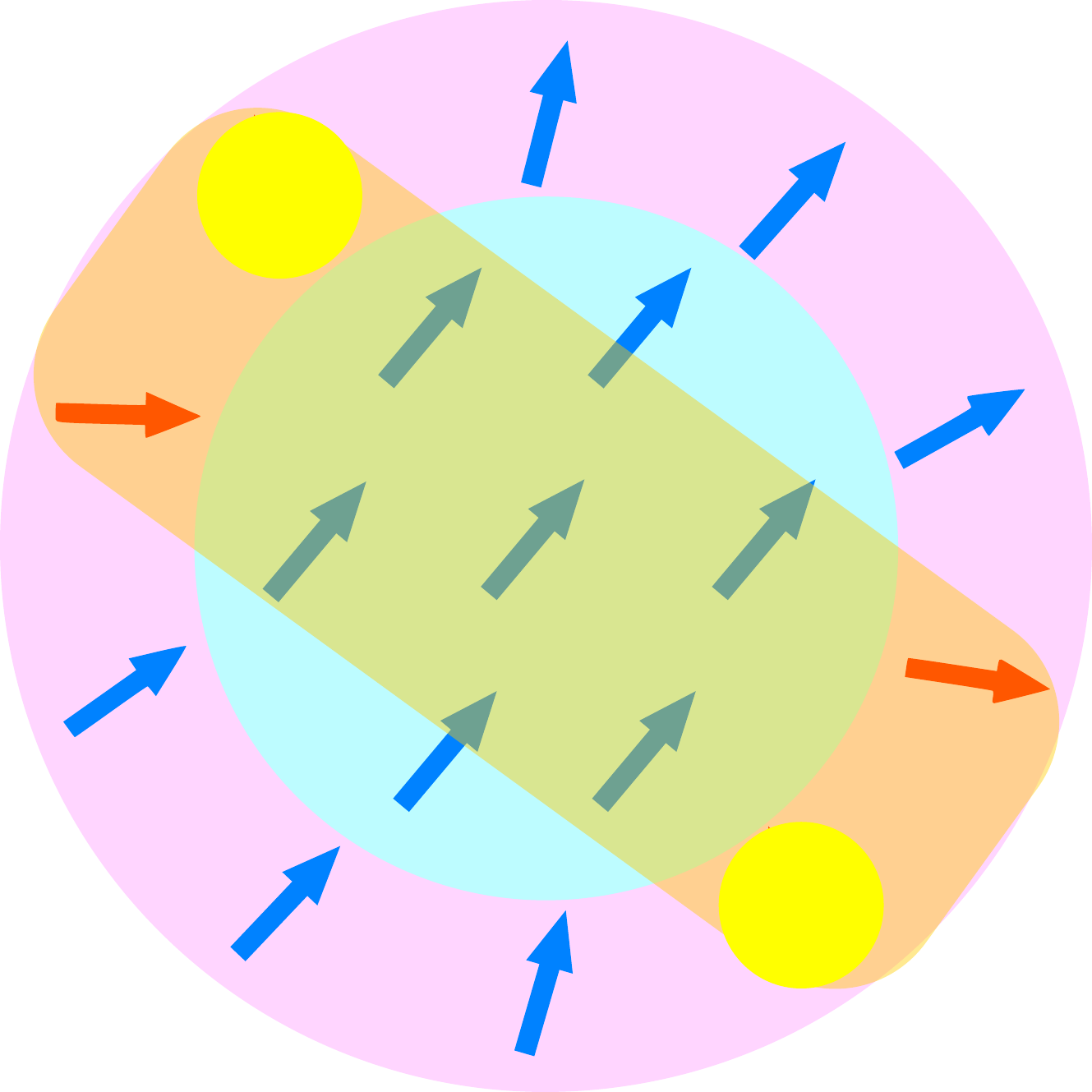}}

\caption{(colour online) (a) Illustration of the spin configuration of one nanoparticle.  The red colour represents the high energy spins.  (b) The position of the magnetic equator depends on the direction of the core magnetization.  (c) The yellow circles represent the random vacancies and the nanoparticle has four spins with higher energy.  (d) The nanoparticle has only two higher energy spins and an effective torque exerted on the nanoparticle magnetization to minimize the overall energy.}\label{fig:EQ}

\end{figure}

The results in Fig.~\ref{fig:EvsK1} clearly show that the distribution of surface vacancies plays a critical role in determining the relationship between the surface energy and the orientation of the core spins. To obtain a qualitative understanding  of what exactly determines the nature and form of this relationship, consider first the magnetic configuration of a spherical nanoparticle at $T\approx 0$ as illustrated schematically in Fig.~\ref{fig:EQ}(a).  Figure~\ref{fig:EQ}(a) shows a nanoparticle with no surface vacancies, a uniformly magnetized core and the surface spins directed inwards at the south pole and outwards at the north pole. While the detailed nature of the magnetic structure at the poles is determined by the minimizing super-exchange interactions and the single ion radial anisotropy at the surface for a given core orientation $\hat\sigma$, it is nevertheless clear from figure Fig.~\ref{fig:EQ}(a) that such a configuration must contain some form of domain wall located at the magnetic equator separating  the two regions surrounding the magnetic poles.  This has been confirmed in simulation studies\cite{pinning} and is shown to be qualitatively similar to a N\'eel domain wall. Because the spins contained within the domain wall are highly frustrated, their energy will be higher than the surface spins located within the regions surrounding the magnetic poles.


Consider now the case shown schematically in Fig.~\ref{fig:EQ}(b) that shows the same system but with the core magnetization rotated\footnote{It is essential to note that when we refer to rotating the spin configuration, the location of the spins themselves do not change. Instead we mean that the individual spins remain fixed at their lattice sites and it is only the direction of their magnetic moment that is rotated}. Qualitatively we would expect that the surface spins rotate  to align with the core spins and that the energy would remain approximately constant. We also note that both spin configurations contain four frustrated spins shown in red. 
Compare this now with the situation shown schematically in Figs.~\ref{fig:EQ}(c) and (d) in which we show essentially the same two spin configurations but with two surface vacancies indicated by the yellow dots. Whereas in the previous example each configuration has four frustrated spins, the spin configuration on the left (Fig.~\ref{fig:EQ}(c)) has four frustrated spins but the spin configuration on the right has only two frustrated spins. Since the frustrated spins (colored red) have a higher energy than the other spins (colored blue) we would expect the energy of the nanoparticle spin configuration on the right to have a lower energy than the spin configuration on the left. 

While the nanoparticles we consider are somewhat more complicated than the very simplified system considered above, the basic counting argument nevertheless provides a qualitative description of the correlation between the vacancy distribution and the surface energy demonstrated in  Fig.~\ref{fig:EvsK1} as a function of the orientation of the core magnetization.  The argument may be summarized as follows: for a given orientation $\hat\sigma$ of the core magnetization, the surface spins may be divided into two distinct regions surrounding the magnetic poles. These regions are separated by a domain wall located on the magnetic equator.  The spins in this region are highly frustrated and have a higher energy than those located in the polar regions. As we rotate the core magnetization the surface spins respond such that the domain wall separating the two polar regions lies in the plane perpendicular to $\hat\sigma$  passing through the origin. As illustrated in Figs.~\ref{fig:EQ}(c) and (d)  the greater number of vacancies located within the domain wall the fewer the number of frustrated spins and the lower the energy of the spin configuration. This is akin to the pinning of domain walls by impurities and defects in standard ferromagnetic materials.

\begin{figure}[t!]

\includegraphics[scale=0.33]{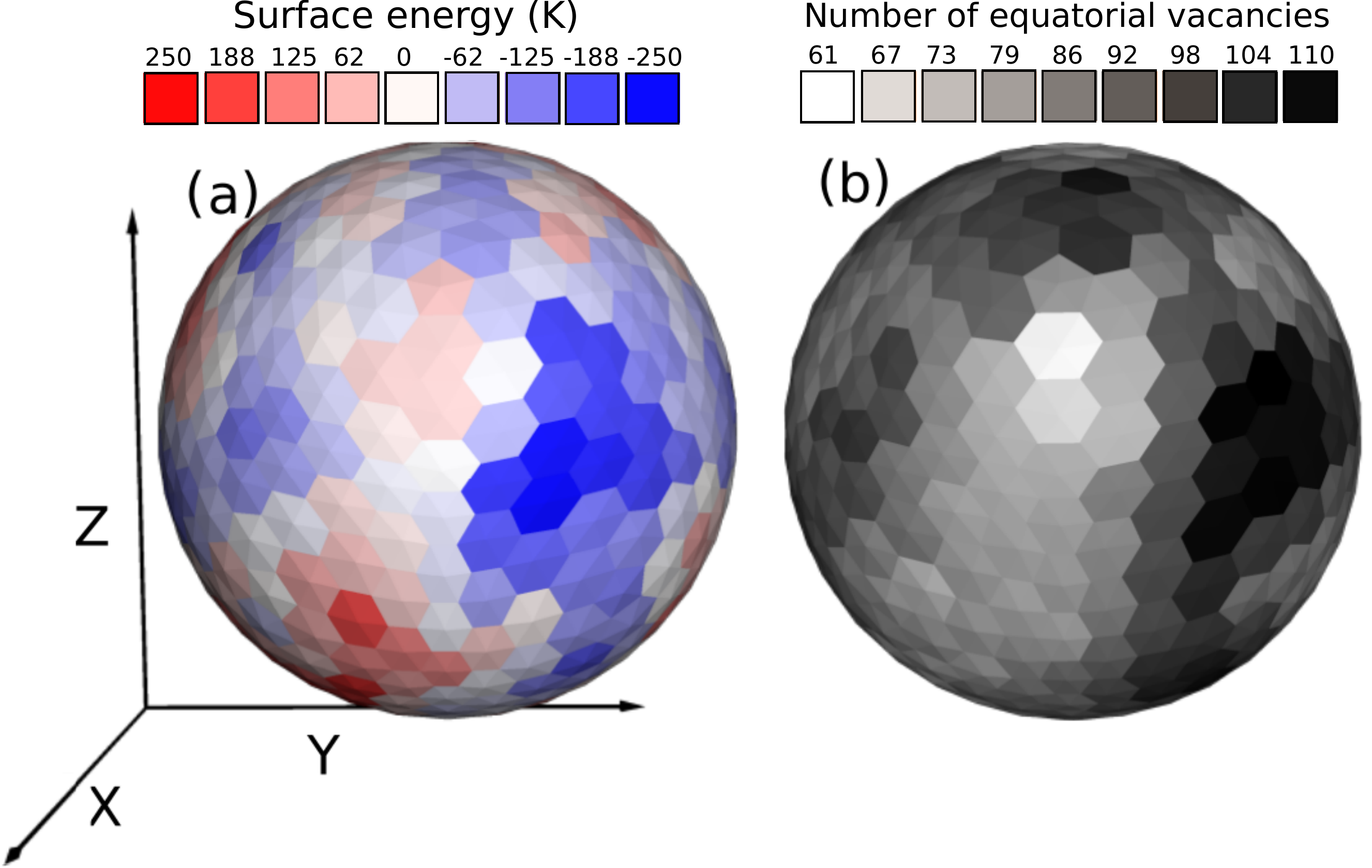}

\caption{(colour online) The surface energy landscape of one nanoparticle with $K_s/k_B$=10~K at (a) T=5 K, and (b) the number of vacancies in the equatorial band associated with each of the mesh points $\{\hat\sigma_i\}$. }\label{fig:Emap1}

\end{figure}

To visualize the dependence of the surface energy of the nanoparticle on the orientation of the core magnetization, we subdivide the convex hull of the points  $\{\hat\sigma_i\}$ on the unit sphere by constructing a Vornoi diagram that consists of a set of polygons (the Vornoi regions) that enclose each of the points $\{\hat\sigma_i\}$. Each Vornoi region is then color coded according to the surface energy (Fig.~\ref{fig:Emap1}~(a)) and the number of vacancies in the equatorial band (Fig.~\ref{fig:Emap1}~(b)) when the core magnetization is in the direction of of $\{\hat\sigma_i\}$. Clearly, the largest numbers of equatorial vacancies coincide with  the directions of the core magnetization which gives the lowest surface energy. In addition, since both the number of vacancies in the equatorial band and the energy of a single nanoparticle given by Eqn.~\eqref{eq:NP-Energy} calculated for a specific orientation $\{\hat\sigma_i\}$ are invariant under the inversion $(\hat\sigma_i \to -\hat\sigma_i)$, 
 the surface energy will consist of a set of local minima comprised entirely of several distinct sets of degenerate pairs. Each pair of local energy minima may be thought of as defining an effective anisotropy axis. The set of such axes will be determined by the specific nature of the surface and will therefore be unique to each nanoparticle and will serve to characterize the effective anisotropy, alluded to earlier, for each of the individual nanoparticles.


\begin{figure}[t!]

\includegraphics[scale=0.25]{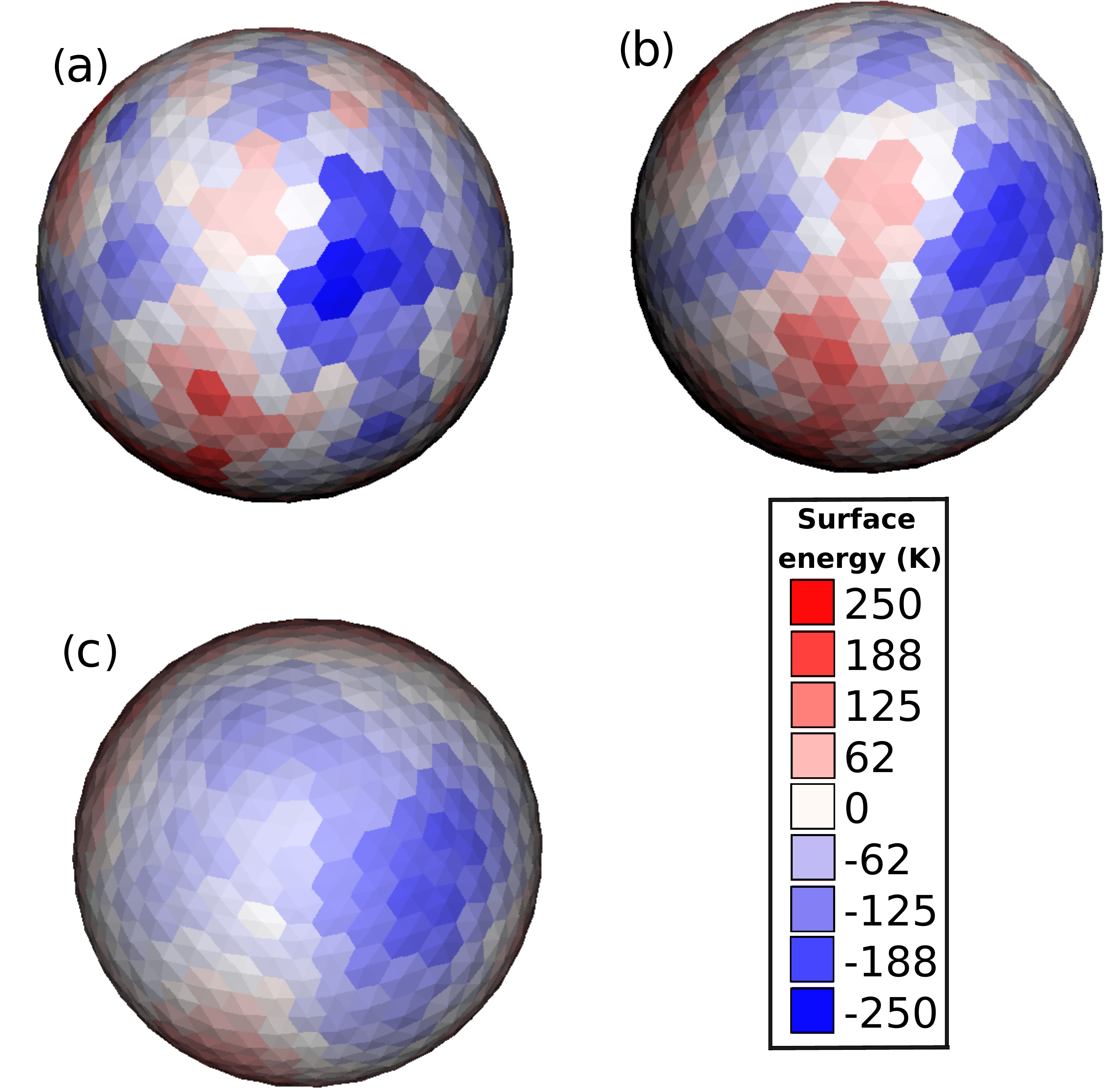}

\caption{(color online) The surface energy landscape of one nanoparticle with $K_s/k_B$=10~K at (a) T=5 K, (b) T=15 K, and (c) T=25 K.  }\label{fig:Emap}

\end{figure}

The energy landscapes for a nanoparticle with $K_s/k_B=10\,\mathrm{K}$ are shown in Fig.~\ref{fig:Emap} for (a) $T=5\,\mathrm{K},\, (b)~ 15\,\mathrm{K}\, ~\mathrm{and}~ (c)~\,25\,\mathrm{K}$. The energy landscapes are characterized by several distinct local extrema and while the overall scale of the variation in the surface energy decreases with increasing temperature, the location of the local extrema are relatively insensitive to temperature. The variation of the surface energy (and hence the effective anisotropy) decreases as the temperature approaches $25~\mathrm{K}$ above which the surface spins disorder. This behaviour is consistent with the discussion presented schematically in Fig.~\ref{fig:EQ} since the spatial distribution of the vacancies (which determines the location of the minima) is independent of temperature, whereas the degree of frustration of the surface spins located in the vicinity of the magnetic equator will decrease as the surface magnetization decreases with increasing temperature (which determines the magnitude of the variation of the surface energy). In addition, above $25~\mathrm{K}$ the surface spins disorder and there is no well defined domain wall. As a consequence, the surface spin configuration has a negligible effect on the dependence of surface energy on the orientation of the core spins. This is consistent with our results in Fig.~\ref{fig:MvsT-NP} where the array magnetization is independent of the surface radial anisotropy at temperatures above 25~K. Furthermore, the increase of the surface effective anisotropy with reducing the temperature (below which the surface spins begin to order) is consistent with experimental observations\cite{surface-disorder} of spherical non-interacting maghemite nanoparticles. Such behavior is expected to be observed in magnetic nanoparticles of materials other than maghemite due to crystalline defects or due to doping with non-magnetic atoms.


In the nanoparticle arrays, the magnetization of each nanoparticle approaches one of the directions that maximize the number of the equatorial vacancies at the surface to reduce the surface energy. At the same time, it tends to maintan an in-plane direction that is close to the magnetizations of the surrounding nanoparticles to minimize the energy of the ferromagnetic dipolar interactions. As a result, the array can be divided into magnetic domains where each group of neighboring nanoparticles tend to have a local magnetization that is  different from other groups. With reducing the radial surface anisotropy, the ferromagnetic long range dipolar interactions become more dominant over the localized surface effective anisotropy and the size of each domain increases. As the size of the magnetic domains increases with reducing the radial anisotropy, the number of the magnetic domains in the array decreases as shown in Figs.~\ref{fig:Ks0T0},\ref{fig:ArrayConfig}.


\section{Conclusions}\label{sec:conclusions}

We have studied a system of spherical maghemite nanoparticles on a two dimensional triangular array using an sLLG approach. As a benchmark, we have first studied dipole-dipole interactions in triangular arrays of classical three-dimensional spins with periodic boundary conditions.  We find, as expected, that triangular arrays of simple dipoles order ferromagnetically below a critical temperature $T_c$=0.663$\mu_0m^2 /4\pi a^3 k_B$, and have an infinitely degenerate ferromagnetic ground state that corresponds to an in-plane net magnetization.  Below the critical temperature, the net magnetization remains in the plane and a six-fold planar anisotropy arises due to an order-from-disorder process at low temperatures.  We have calculated the anisotropy barrier and we find excellent agreement with the simulations.

Simulations of a triangular array of 7.5~nm diameter maghemite nanoparticles were performed using the point dipole approach. The temperature dependence of the magnetization for array of nanoparticles that have no surface anisotropy may be mapped onto the corresponding curve for simple dipole array by a simple re-scaling of the magnetization and temperature. For particles with a surface radial anisotropy, an {\it effective} random temperature-dependent anisotropy arises which competes with the dipole interactions and leads to a reduction of the low temperature magnetization of the array.  Although the magnetization of each nanoparticle increases with decreasing temperature, the {\it effective} anisotropy increases more rapidly, so as to decrease the alignment between the nanoparticle magnetizations in the array when $K_s/k_B$$\geq$5~K.  We find that the radial anisotropy results in the formation of magnetic domains in the array where the number of the domains increases with $K_s$.

For individual nanoparticles, the competition between the radial anisotropy and the super-exchange interactions on the surface results in a N\'{e}el-like domain wall on the magnetic equator.  The high energy domain wall on the magnetic equator gives rise to an inhomogeneity in the energy of the surface spins that depends on the direction of the nanoparticle magnetic moment.  The random vacancy distribution on the octahedral sites results in a static inhomogeneity in the distribution of surface spins.  The interplay between the  energy distribution on the surface and the static (inhomogeneous) spin distribution gives rise to an effective torque that is exerted on the nanoparticle magnetization to minimize the energy of the surface spins.  In other words, each nanoparticle has a unique {\it effective} anisotropy that is temperature dependent. We find that the effective anisotropy increases with the radial anisotropy constant $K_s$ and with reducing the temperature.

 Finally, as mentioned in the introduction, the nature of the equilibrium magnetization in an order-disorder transition depends on the specific nature of the disorder. It has been shown that competing forms of disorder, most notably structural and thermal fluctuations, can give rise to phase transitions that reflect the competing nature of the two forms of disorder. It is interesting therefore to speculate, given that the thermal disorder decreases with decreasing temperature while the disorder due to the effective anisotropy increases with decreasing temperature, the competition between these two forms of disorder might also result in some form of transition between two distinct magnetic states. Such a possibility may account for the presence of the domains observed in the equilibrium configurations shown in Fig.~\ref{fig:ArrayConfig} in which the spins exhibit a significant out of plane component in comparison with Fig.~\ref{fig:Ks0T0}.
\\

\acknowledgments

This work was supported by the Natural Sciences and Engineering Research Council (NSERC) of Canada (RGPIN-2018-0501), Compute Canada, and the University of Manitoba.

\bibliography{g-Fe2O3NPs_triarray4}

\end{document}